\documentclass[conference]{IEEEtran}
\IEEEoverridecommandlockouts

\usepackage{cite}
\usepackage{amsmath,amssymb,amsfonts}
\usepackage{algorithmic}
\usepackage{graphicx}
\usepackage{CJKutf8}
\usepackage{stfloats}
\usepackage{textcomp}
\usepackage{xcolor,color,url}
\usepackage{bm}
\usepackage{booktabs}
\usepackage{array}

\begin{document}
\begin{CJK}{UTF8}{gbsn}
\title{A Lightweight and Real-Time Binaural Speech Enhancement Model with Spatial Cues Preservation\\
\thanks{This work is supported by the National Natural Science Foundation of China (62101523), Anhui Province Major Science and Technology Research and Development Project (S2023Z20004) and USTC Research Funds of the Double First-Class Initiative (YD2100002008). 
}}
\author{Jingyuan Wang$^1$, Jie Zhang$^1$, Shihao Chen$^1$, Miao Sun$^2$\\
$^1$NERC-SLIP, University of Science and Technology of China (USTC), Hefei, China\\
$^2$School of Information and Communication Engineering, Guangzhou Maritime University, Guangzhou, China\\
jywg@mail.ustc.edu.cn; jzhang6@ustc.edu.cn; shchen16@mail.ustc.edu.cn; m.sun.1@hotmail.com}

\maketitle

\begin{abstract}
Binaural speech enhancement (BSE) aims to jointly improve the speech quality and intelligibility of noisy signals received by hearing devices and preserve the spatial cues of the target  for natural listening. Existing methods often suffer from the compromise between noise reduction (NR) capacity and spatial cues preservation (SCP) accuracy and a high computational demand in complex acoustic scenes. In this work, we present a learning-based lightweight binaural complex convolutional network (LBCCN), which excels in NR by filtering low-frequency bands and keeping the rest. Additionally, our approach explicitly incorporates the estimation of interchannel relative acoustic transfer function to ensure the spatial cues fidelity and speech clarity. Results show that the proposed LBCCN can achieve a comparable NR performance to state-of-the-art methods under fixed-speaker conditions, but with a much lower computational cost and a certain degree of SCP capability. The reproducible code and audio examples are available at
\url{https://github.com/jywanng/LBCCN}.
\end{abstract}


\section{Introduction}
Speech enhancement (SE) aims to improve the speech quality and intelligibility by reducing background noise, including single-channel and multichannel algorithms, which have developed rapidly over past few decades~\cite{intro1}. For binaural listening devices, e.g., hearing aid (HA), headphone, cochlear implant, which can be used to improve the listening level of hearing-impaired persons, it is expected to not only increase the speech clarity of noisy recordings but also perceive the stereo acoustic scene from the enhanced signals (i.e., spatial awareness). The latter requirement is closely related to binaural cues, which are essential for sound localization~\cite{intro2}. In this sense, conventional SE methods (with a single audio output) usually cannot be directly applied in the binaural context. {\it The focus of this work is thus on the binaural SE (BSE) with a joint noise reduction (NR) and spatial cues preservation (SCP) of the target speaker}.

Often-used binaural cues include interaural phase difference (IPD), interaural time difference, interaural level difference (ILD), magnitude squared coherence~\cite{intro2, intro3, intro4}. Statistical signal processing was initially applied to the BSE task, e.g., minimum variance distortionless response beamformer~\cite{minimum1}, linearly-constrained minimum variance beamformer~\cite{linearly1,linearly2,zhang2},  multichannel Wiener filter~\cite{wiener1,wiener2,wiener3,zhang3}, parametric unconstrained beamformer~\cite{zhang1}, by designing two filters at the ears in order to achieve the stereo outputs. These well-established methods can work very well and fast in stationary acoustic conditions, while the efficacy in non-stationary cases is rather limited.

Deep neural network (DNN) is capable of learning a non-linear mapping function from the provided data to the target, which has been widely applied to both single-channel and multichannel SE, even with a better NR performance in non-stationary conditions~\cite{se1, se2, se3, se4}. Similarly for BSE, several learning-based methods have been proposed. For instance, in \cite{bse1} a complex-valued DNN was proposed to suppress interference and preserve the target binaural cues. However, the use of real-valued prediction heads limits the ability to fully exploit the inherent advantages of complex-valued representations. The short-time objective intelligibility (STOI)-optimal masking was adopted in~\cite{bse2}, which can enhance the signal-to-noise ratio (SNR) by separately processing the binaural channels, but without a guarantee on the preservation of spatial cues. Further, in~\cite{bse3} two Conv-TasNet~\cite{se3} networks were configured for binaural channels, which is more promising in performance, while the whole model is computationally intensive and requires substantial memory resources. More recently, the state-of-the-art (SOTA) BSE results can be found in~\cite{bse4}, which exploits a complex-valued transformer, yet with a very large model size and high complexity.

As BSE algorithms have to be deployed on latency-sensitive and low-resource listening devices, e.g., HAs, the computationally cheap and low-latency (real-time) models are thus more preferable. To achieve this, we propose a \textbf{l}ightweight \textbf{b}inaural \textbf{c}omplex \textbf{c}onvolutional \textbf{n}etwork (LBCCN) for joint binaural NR and SCP. As usually low-frequency signal components can enhance the segregation of competing voices, leading to a better speech understanding in noise~\cite{band1,band2,band3}, the proposed model selectively filters low-frequency spectrum and keeps the remaining frequencies unchanged. It is shown that this operation can even improve the speech intelligibility at a small sacrifice in the speech quality, but more importantly heavily reduces the computational complexity. 
The proposed LBCCN leverages pure convolutional network and interchannel relative acoustic transfer function (RATF)-based predictors instead of direct masking~\cite{bse1, bse4} to improve the SCP and reduce model parameters. Experimental results on a synthesized dataset demonstrate the superiority of the proposed model in aspects of the BSE performance and computational cost.

\section{Proposed LBCCN Method}
\begin{figure*}[ht] 
\centering 
\includegraphics[scale=0.45]{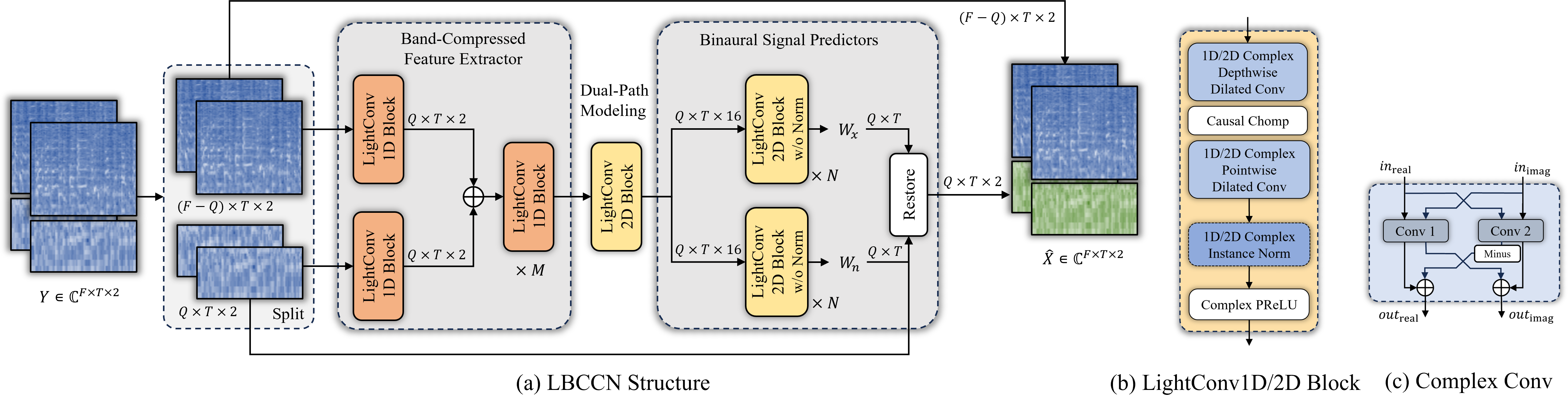} 
\caption{The proposed LBCCN BSE model, which mainly consists of band-compressed feature extractor (operates on the lower-$Q$ frequency bands), dual-path modeling and signal predictors. The LightConv1D operates on the frequency dimension, and LightConv2D on both time and frequency dimensions.} 
\label{fig1} 
\end{figure*}

In the time domain, the binaural signals received by the two ears can be written as
\begin{equation}
\begin{aligned}
\bm{y}_i = \bm{x}_i + \bm{n}_i = \bm{h}_i \ast \bm{s} +  \bm{n}_i, \quad i\in\{L,R\},
\end{aligned}\label{eq_1}
\end{equation}
where $\bm{s}$ represents the original speech signal of interest, $\bm{x}_i$ the recorded signal component, $\bm{n}_i$  the diffuse
isotropic noise component\footnote{The spherically diffuse isotropic noise field was shown to be a reasonable approximation of daily practical noise fields in, e.g., an office or car \cite{isotropic1}. Given a noise source and HRIRs of all directions, the diffuse isotropic noise component in this work is synthesized by convolving the noise source and each HRIR and then averaging over all convolutions.}, $\bm{h}_i$ the head-related impulse responses (HRIRs) of the target speech source with respect to the two ears,  $\ast$ the linear convolution, respectively. 
{\it The goal of the considered BSE is to extract the target signal components $\bm{x}_i,i\in\{L,R\}$ from the received measurements $\bm{y}_i$ and preserve the corresponding spatial cues.}
The proposed LBCCN model mainly consists of the band-compressed feature extractor, dual-path modeling and binaural signal predictors, which is depicted in Fig.~\ref{fig1}. Next, we will introduce each module in detail.

\subsection{Band-Compressed Feature Extractor}
In order to reduce the computational complexity, we selectively enhance some frequency bands and retain the recorded components in the remaining bands. Our focus is on filtering low-frequency bands, say $Q$ bands, because people perceive the speech signals depending more on low-frequency components, e.g., the fundamental components of vowels, consonants, which are more useful for the segregation of competing voices and speech understanding in noise~\cite{band1,band2,band3}.

The proposed LBCCN fisrt applies the short-time Fourier transform (STFT) to the input noisy speech signals to obtain time-frequency (TF) representations, followed by two 1D lightweight convolutional (LightConv1D) blocks to compress the frequency spectrum.  The LightConv1D blocks are used to process the selected and unselected bands on the frequency dimension, respectively, and their dimensions should adapt to the respective frequency-band numbers. As shown in Fig.~\ref{fig1}(b), each LightConv1D block employs a depth-wise separable causal convolution layer (with depthwise and pointwise convolutions and causal chomp) to ensure the low-complexity and causality. To achieve a larger receptive field, we employ dilated convolutions in these blocks. All operations are complex-valued, such as the complex convolutional layer in Fig.~\ref{fig1}(c). These layers are followed by instance normalization and PReLU activation functions. The outputs of these blocks are combined and then processed by $M$ additional LightConv1D blocks for feature extraction. As our focus remains on the low-frequency bands, the compressed latent space dimension is tailored to match that of the low-frequency bands.

\subsection{Dual-Path Modeling}
In order to leverage both temporal and frequency speech features for BSE, we employ a 2D lightweight convolutional (LightConv2D) block instead of traditional long short-term memory (LSTM) network~\cite{dualpath1} or Transformer~\cite{dualpath2} for dual-path modeling. This design can also reduce the model complexity. The structure of the LightConv2D block is similar to LightConv1D in Fig.~\ref{fig1}(b).

\subsection{Binaural Signal Predictors}
Existing BSE models usually directly apply estimated masks to predict the binaural outputs~\cite{bse1, bse4}, given by
\begin{equation}
\begin{aligned}
&\hat{X}_i = M_i \odot Y_i, \quad i\in\{L,R\},
\end{aligned}\label{eq_2}
\end{equation}
where \( Y_i \) and \( M_i \) are the noisy STFT coefficients and complex ideal ratio mask (cIRM) at the two ears, $\odot$ the Hadamard product, respectively. The masks can be seen as the outputs of the LightConv2D blocks in the binaural signal predictors in Fig.~1 and applied to the noisy inputs to recover the target signals via inverse STFT (iSTFT). Hence, the direct masking is a special case of the proposed LBCCN model, referred to as \textbf{LBCCN (Masks)} in Section III-B for comparison.

In~\cite{predictor1, predictor2}, cascaded binaural speech separation methods were proposed, which first predict the mask for one channel and estimate the RATF. The estimated RATF is then used to calculate the output of the other channel. The inclusion of the RATF estimation is concerned with the preservation of the target binaural cues. It is clear that this cascaded pipeline suffers from the error accumulation, as the overall signal clarity depends on the RATF accuracy. To show this, we combine the proposed LBCNN backbone with the masking and RATF estimation modules for comparison in Section III-B, referred to as \textbf{LBCCN (Mask+RATF)}.

As the binaural cues are closely related to the RATF,  in this work we regard the RATF of the target as an implicit prediction target for a better SCP, see the binaural signal predictors in Fig.~\ref{fig1}(a). Letting $H^x_L$ and $H^x_R$ denote the ATF of the target source with respect to the two ears, and $H^n_L$ and $H^n_R$ are the average ATF of the noise source in the diffuse isotropic noise field, respectively, we define the frequency-dependent RATFs as $W_x=H^x_L/H^x_R$ and $W_n=H^n_L/H^n_R$. The clean binaural  signal components at each TF bin $(t,f)$ can be written as
\begin{align}
X_L = W_x X_R, \quad
X_R = \frac{Y_L - W_n Y_R}{W_x - W_n}, \label{eq:clean_LR}
\end{align}
by reformulating the STFT-domain signal model.
Unlike the cIRMs \( M_L \) and \( M_R \), \( W_x \) and \( W_n \) have a determined physical meaning. Predicting \( W_x \) and \( W_n \) directly reflects the SCP and the required number of neurons can be largely reduced, as the number of binaural microphones is usually very small.

We use two prediction heads to estimate the RATFs \( W_x \) and \( W_n \) in the TF domain, which include \(N\) LightConv2D blocks. To alleviate the impact of the amplitude variation on the training stability, we remove the normalization layers in the LightConv2D blocks. After obtaining \( W_x \) and \( W_n \), we follow \eqref{eq:clean_LR}  to {\it Restore} the STFT-domain signal components.

\subsection{Loss Function}

Let \(\hat{\bm{x}}\) represent the predicted speech signal, the predicted noise signal \(\hat{\bm{n}}\) is then given by

\begin{equation}
\begin{gathered}
\hat{\bm{n}} = \bm{x} + \bm{n} - \hat{\bm{x}}.
\end{gathered}\label{eq_4}
\end{equation}
We adopt the weighted signal and noise losses to construct the overall loss function for model training as
\begin{equation}
\begin{gathered}
\mathcal{L}_{\text{total}} = k\mathcal{L}(\hat{\bm{x}}, \bm{x}) + (1-k)\mathcal{L}(\hat{\bm{n}}, \bm{n}),
\end{gathered}\label{eq_5}
\end{equation}
where \(\mathcal{L}\) is defined similarly as that in~\cite{bse4}:
\begin{equation}
\begin{gathered}
\mathcal{L} =  \mathcal{L}_{\text{SNR}} + 10 \mathcal{L}_{\text{STOI}} +  \mathcal{L}_{\text{IPD}} + 10 \mathcal{L}_{\text{ILD}},
\end{gathered}\label{eq:loss}
\end{equation}
which however is calculated over the selected bands. The components in (\ref{eq:loss}) are computed as:
\begin{equation}
\begin{aligned}\nonumber
\mathcal{L}_{\text{SNR}} &= -\frac{1}{2}\sum_{i\in\{L,R\}}{10\log_{10}{\left(\frac{\Vert \bm{x}_i \Vert^2}{\Vert \bm{\hat{x}}_i - \bm{x}_i \Vert^2}\right)}},\\
\mathcal{L}_{\text{STOI}} &= -\frac{1}{2}\sum_{i\in\{L,R\}} \text{STOI}\left(\bm{x}_i, \bm{\hat{x}}_i\right),\\
\mathcal{L}_{\text{ILD}} &= \frac{20}{TF}\sum_t\sum_f\left(\log_{10}{\left(\frac{\lvert X_L \rvert}{\lvert X_R \rvert}\right)} - \log_{10}{\left(\frac{\lvert \hat{X}_L \rvert}{\lvert \hat{X}_R \rvert}\right)}\right),\\
\mathcal{L}_{\text{IPD}} &= \frac{1}{TF}\sum_t\sum_f\left(\arctan{\left(\frac{\lvert X_L \rvert}{\lvert X_R \rvert}\right)} - \arctan{\left(\frac{\lvert \hat{X}_L \rvert}{\lvert \hat{X}_R \rvert}\right)}\right),
\end{aligned}
\end{equation}
where \(\text{STOI}(\cdot)\) represent the short-time objective intelligibility (STOI) measure~\cite{evaluate1}, \(TF\) is the total number of TF bins, i.e., the ILD and IPD losses are averaged over all TF bins.

\begin{table*}[htbp]

    \centering
    \caption{Performance Comparison of BSE Methods in terms of the SNR Level.}
    \renewcommand{\arraystretch}{0.9}
    \resizebox{0.9\textwidth}{!}{ 
    \begin{tabular}{c|cccc|cccc|cccc}
        \toprule
        \multicolumn{1}{c|}{Input SNR} & \multicolumn{4}{c|}{-10 dB} & \multicolumn{4}{c|}{-5 dB} & \multicolumn{4}{c}{0 dB} \\
        \midrule
        Method & MBSTOI $\uparrow$ & $\Delta$PESQ $\uparrow$ & \(\mathcal{L}_{\text{ILD}}\) $\downarrow$ & \(\mathcal{L}_{\text{IPD}}\) $\downarrow$ & MBSTOI $\uparrow$ & $\Delta$PESQ $\uparrow$ & \(\mathcal{L}_{\text{ILD}}\) $\downarrow$ & \(\mathcal{L}_{\text{IPD}}\) $\downarrow$ & MBSTOI $\uparrow$ & $\Delta$PESQ $\uparrow$ & \(\mathcal{L}_{\text{ILD}}\) $\downarrow$ & \(\mathcal{L}_{\text{IPD}}\) $\downarrow$ \\
        \midrule
        DBSEnh~\cite{bse1} & 0.72 & 0.01 & 4.84 & 0.94 & 0.80 & 0.05 & 4.47 & 0.80 & 0.87 & 0.10 & 4.04 & 0.59 \\
        BiTasNet~\cite{bse3} & 0.80 & 0.35 & 5.16 & 1.05 & 0.86 & 0.52 & 4.53 & 0.91 & 0.90 & 0.71 & 3.90 & 0.80 \\
        BCCTN~\cite{bse4} & 0.81 & \textbf{0.41} & \textbf{3.55} & 0.83 & 0.87 & \textbf{0.66} & \textbf{2.99} & 0.70 & 0.92 & \textbf{0.97} & \textbf{2.10} & 0.49 \\
        LBCCN & \textbf{0.88} & 0.34 & 3.64 & \textbf{0.68} & \textbf{0.91} & 0.61 & 3.14 & \textbf{0.59} & \textbf{0.94} & 0.87 & 2.33 & \textbf{0.47} \\
        \midrule
        \multicolumn{1}{c|}{Input SNR} & \multicolumn{4}{c|}{5 dB} & \multicolumn{4}{c}{10 dB} & \multicolumn{4}{|c}{Average} \\
        \midrule
        Method & MBSTOI $\uparrow$ & $\Delta$PESQ $\uparrow$ & \(\mathcal{L}_{\text{ILD}}\) $\downarrow$ & \(\mathcal{L}_{\text{IPD}}\) $\downarrow$ & MBSTOI $\uparrow$ & $\Delta$PESQ $\uparrow$ & \(\mathcal{L}_{\text{ILD}}\) $\downarrow$ & \(\mathcal{L}_{\text{IPD}}\) $\downarrow$ & MBSTOI $\uparrow$ & $\Delta$PESQ $\uparrow$ & \(\mathcal{L}_{\text{ILD}}\) $\downarrow$ & \(\mathcal{L}_{\text{IPD}}\) $\downarrow$ \\
        \midrule
        DBSEnh~\cite{bse1} & 0.91 & 0.08 & 3.99 & 0.50 & 0.93 & 0.10 & 3.98 & 0.41 & 0.85 & 0.06 & 4.22 & 0.64 \\
        BiTasNet~\cite{bse3} & 0.93 & 0.73 & 3.58 & 0.72 & 0.94 & 0.72 & 3.38 & 0.68 & 0.89 & 0.61 & 4.09 & 0.83 \\
        BCCTN~\cite{bse4} & 0.95 & \textbf{1.03} & \textbf{1.63} & \textbf{0.37} & 0.96 & \textbf{1.07} & \textbf{1.32} & \textbf{0.28} & 0.90 & \textbf{0.86} & \textbf{2.27} & 0.58 \\
        LBCCN & \textbf{0.95} & 1.01 & 1.97 & 0.41 & \textbf{0.96} & 1.01 & 1.78 & 0.37 & \textbf{0.93} & 0.78 & 2.53 & \textbf{0.50} \\
        \bottomrule
    \end{tabular}
    }
    \label{tab1}
\end{table*}

\begin{table}[htbp]
    \centering
    \renewcommand{\arraystretch}{0.9}
    \caption{The Model Complexity, Computational Requirements and Real Time Factor (RTF) of Different BSE Methods.}
    \scriptsize
    \begin{tabular}{c|cccc}
        \toprule
        Method & Parameters $\downarrow$ & MACs $\downarrow$ & RTF $\downarrow$\\
        \midrule
        DBSEnh~\cite{bse1} & 10.5 M & 11.1 G & \textbf{0.022}\\
        BiTasNet~\cite{bse3} & 1.7 M & 17.2 G & 0.329\\
        BCCTN~\cite{bse4} & 11.1 M & 12.7 G & 0.228\\
        LBCCN & \textbf{38.0 K} & \textbf{216.3 M} & 0.054\\
        \bottomrule
    \end{tabular}
    \label{tab2}
\end{table}

\begin{table}[htbp]
    \centering
    \caption{Ablation Study on Band Compression and  Signal Predictors.}
    \renewcommand{\arraystretch}{0.9}
    \scriptsize 
    \resizebox{0.48\textwidth}{!}{
    \begin{tabular}{c|ccccc}
        \toprule
        Method & MBSTOI $\uparrow$ & PESQ $\uparrow$ & \(\mathcal{L}_{\text{ILD}}\) $\downarrow$ &\(\mathcal{L}_{\text{IPD}}\) $\downarrow$  & RTF $\downarrow$\\
        \midrule
        LBCCN (\(Q=30\)) & 0.93 & 0.64 & \textbf{2.41} & \textbf{0.47} & 0.054\\
        LBCCN (\(Q=40\)) & \textbf{0.93} & \textbf{0.78} & 2.53 & 0.50 & \textbf{0.054}\\
        LBCCN (\(Q=64\)) & 0.93 & 0.69 & 2.96 & 0.72 & 0.057\\
        LBCCN (\(Q=129\)) & 0.92 & 0.75 & 3.48 & 1.07 & 0.063\\
        \midrule
        LBCCN (Masks) & 0.86 & 0.17 & 2.54 & 0.51 & 0.054\\
        LBCCN (Mask+RATF) & 0.70 & 0.14 & \textbf{2.44} & \textbf{0.49} & 0.054 \\
        \textbf{LBCCN (RATFs)} & \textbf{0.93} & \textbf{0.78} & 2.53 & 0.50 & \textbf{0.054}\\
        \bottomrule
    \end{tabular}}
    \label{tab3}
\end{table}

\section{Experiments}
In this section, we will explain the implementation of the proposed LBCCN approach together with the experimental comparison with SOTA models.
\subsection{Experimental Setup}
\textbf{Dataset:} 
To verify the efficacy of the proposed LBCCN model, we create a spatialized anechoic audio dataset, where the clean speech source originates from Librispeech~\cite{dataset1} and noise source from NoiseX-92~\cite{dataset2}. The HRIRs are derived from the CIPIC~\cite{dataset3} dataset, which contains 45 subjects, each measured at 25 azimuth angles and 50 elevations. Data from 36 subjects are used for training and validation, and the remaining 9 unseen subjects for testing. Clean speech signals are convolved with HRIRs from a fixed direction (45 degrees to the right of center) to generate the binaural clean signal components, and noise signals are convolved with the HRIRs but averaged over all directions.
The signal and noise components are added together at a random SNR ranging from -10 dB to 10 dB, resulting in 50000 2-seconds samples ($\approx$28 hours in total). The samples are divided into at a ratio of 8:1:1 for training, testing and validation, and all are downsampled to 16 kHz.

\textbf{Training details:} 
We apply a 256-point STFT to the noisy binaural signals using a Hanning window with a hop size of 128. The lowest 40 frequency points are selected for BSE, i.e., \(Q=40\). The weight \(k\) in \eqref{eq_5} is set to 0.5. The feature extractor has $M$ = 2 LightConv1D blocks, and the signal predictor has $N$ = 3 LightConv2D blocks.
The LightConv blocks have output channels of \{40, 40, 40, 40\}, \{16\} and \{16, 16, 1\}, kernel sizes of 5, 9 and 9, and dilation rates of \{1, 1, 2, 4\}, \{1\} and \{1, 2, 4\} for the band-compressed feature extractor, dual-path modeling and binaural signal predictor, respectively. Padding is applied in each module to keep the output dimensions consistent. 
The Adam optimizer~\cite{setup1} is used for the training of LBCCN with an initial learning rate of 0.0001, and the model is trained for 20-50 epochs until convergence.

\textbf{Evaluation metrics:} 
For the comparison of BSE models, we use the modified binaural STOI (MBSTOI)~\cite{evaluate2} and the gain of perceptual evaluation of speech quality ($\Delta$PESQ)~\cite{evaluate3} to measure the NR performance (the higher, the better). The losses \(\mathcal{L}_{\text{ILD}}\) and \(\mathcal{L}_{\text{IPD}}\) (averaged over all TF bins) in Section II-D are used to measure the SCP errors (the lower, the better).

\subsection{Experimental Results}

First, we compare the performance of the proposed LBCCN with DBSEnh~\cite{bse1}, BiTasNet~\cite{bse3} BCCTN~\cite{bse4}. The comparison models are reproduced strictly according to the published papers. The obtained results are summarized in Table~\ref{tab1}.  Our method achieves the highest speech intelligibility in MBSTOI in all SNR conditions, which is more obvious in more noisy conditions. Meanwhile, the SCP errors of the proposed method  are competitive with the BCCTN model, which are lower than the other two methods. It seems that filtering on low-frequency bands (e.g., LBCCN) is more helpful for the preservation of IPD.
In Table~\ref{tab2}, we compare the model size and computational complexity of these approaches, which are calculated using the \texttt{thop} package\footnote{\url{https://github.com/Lyken17/pytorch-OpCounter}.}. Clearly, the LBCCN requires significantly less parameters and MACs compared to other methods. Also, the real-time factor (RTF)\footnote{The RTF is measured with Intel(R) Xeon(R) E5-2620 v4 CPU.} of LBCCN is much smaller than that of BiTasNet and BCCTN and slightly higher than that of DBSEnh. Note that the proposed LBCCN has an obvious superiority in performance over DBSEnh.

\begin{figure}[ht] 
\centering 
\includegraphics[scale=0.47]{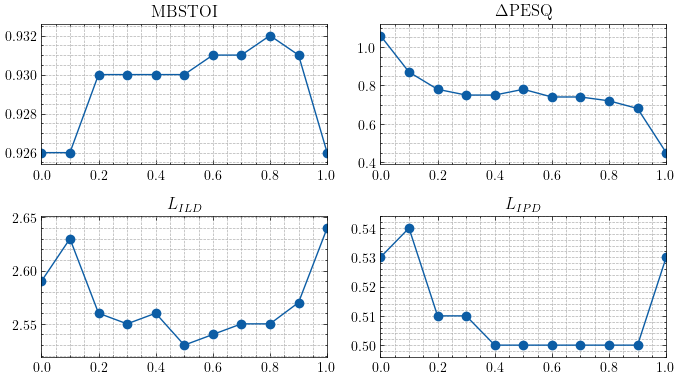} 
\caption{The impact of the signal and noise losses weight \(k\) in (5).} 
\label{fig2} 
\end{figure}

Second, we conduct ablation studies on the frequency band selection and signal predictor in Table~\ref{tab3}. It shows that filtering on the $Q$ selected frequency bands has an impact on both the performance and RTF. In general, the case of $Q$ = 40 (i.e., the upper-bound frequency equals 8$\times$40/129 = 2.48kHz) returns the best NR performance, which almost covers all important bands for the estimation of speech intelligibility~\cite{ansi1997methods}, and reducing the selected bands to $Q$ = 30 can improve the SCP a bit. On the other side, the RTF increases in terms of $Q$. Therefore, we choose $Q$ = 40 unless stated elsewhere.
In addition, the use of signal predictors does not impact the RTF (as the model structures are the same), but influences the performance. The \textbf{LBCCN (Masks)} that generates two masks obtains the worst performance. The \textbf{LBCCN (Mask+RATF)} that generates a mask and an RATF achieves a better SCP, since estimating a single RATF is easier than estimating two by \textbf{LBCCN (RATFs)}. The proposed \textbf{LBCCN (RATFs)} that directly outputs two RATFs can significantly increase the NR capacity at the tiny loss in SCP. 

Finally, we analyze the impact of the  weight parameter \(k\)  in \eqref{eq_5} in Fig.~\ref{fig2}, where the performance is averaged over all noise conditions. We observe that the choice of $k$ = 0.5 yields the best trade-off between the NR performance and SCP errors, since the average SNR of the considered training set is 0dB. That is, the best balance weight $k$ should depend on the average SNR level of the training samples.

\section{Conclusion}
In this paper, we proposed a new BSE approach for joint binaural NR and SCP, called LBCCN, which is  a lightweight, low-complexity and real-time model. The model complexity was reduced by selectively operating on low-frequency bands that are more related to speech understanding in noise and applying the lightweight convolutional blocks to replace normal convolutions. The binaural cues can be better preserved owing to the  inclusion of an explicit estimation of the target RATF. It was shown that the proposed LBCCN can achieve the best performance at most SNR levels, but a more promising superiority in complexity.

\end{CJK}
\end{document}